\documentclass[twocolumn,english,aps,prl,superscriptaddress]{revtex4}
\usepackage{amsmath}
\usepackage{graphicx}
\usepackage{snapshot}
\makeatletter

\usepackage{amsfonts}
\usepackage{amsthm}
\usepackage{mathrsfs}
\usepackage{graphics}
\usepackage{lscape}
\usepackage{changebar}
\usepackage{subfigure}
\usepackage{epsfig}
\usepackage{babel}
\usepackage{array}
\usepackage{color}
\usepackage{cmap}
\usepackage{graphicx}
\usepackage{amsmath}
\usepackage{mathtools}
\usepackage[T1]{fontenc}
\newcommand{\edit}[1]{{\color{black}{#1}}}

\makeatother

\begin{document}

\title{General scaling relations for locomotion in granular media}

\author{James Slonaker$^1$, D. Carrington Motley$^1$, Qiong Zhang$^1$, Stephen Townsend$^1$, Carmine Senatore$^1$, Karl Iagnemma$^1$, Ken Kamrin}
 \email{kkamrin@mit.edu}
 \affiliation{Department of Mechanical Engineering, MIT, Cambridge, MA 02139, USA}

\date{\today}

\begin{abstract}

{We derive a general dimensionless form for granular locomotion, which is validated in experiments and Discrete Element Method (DEM) simulations.  The form instructs how to scale size, mass, and driving parameters in order to relate dynamic behaviors of different locomotors in the same granular media.  The scaling can be derived by assuming intrusion forces arise from Resistive Force Theory (RFT) or equivalently by assuming the granular material behaves as a continuum obeying a frictional yield criterion.   The scalings are experimentally confirmed using pairs of wheels of various shapes and sizes under many driving conditions in a common sand bed. We discuss why the two models provide such a robust set of scaling laws even though they neglect a number of the complexities of granular rheology.  Motivated by potential extra-planetary applications, the dimensionless form also implies a way to predict wheel performance in one ambient gravity based on tests in a different ambient gravity. We confirm this using DEM simulations, which show that scaling relations are satisfied over an array of driving modes even when gravity differs between scaled tests.}

\end{abstract}

\pacs{PACS number(s): }

\maketitle

\section{Introduction}
Due to the complexity of the constitutive behavior of granular media \cite{goddard2014continuum} dynamic interactions between grains and solid bodies are challenging to model without resorting to grain-by-grain discrete particle methods.  When materials have a simple constitutive behavior, one benefit is the ability to develop straightforward scaling laws, which can be used to study material dynamics in complicated geometries. Newtonian fluids, for example, have a well-known rheology governed by two material parameters, viscosity and density, which leads to a handful of well-known dimensionless parameters (e.g. Reynolds number, Froude number) that can be used for controlled scaling analyses.  In contrast, granular media are more complex and the constitutive behavior is still widely debated.  

Granular materials display a diversity of phenomena that require special constitutive treatment to model, such as history- and preparation-dependent strengthening and dilation \cite{schofield1968critical,wood1990soil,roux1998texture}, flow anisotropy and normal stress differences \cite{depken07,weinhart2013coarse,luding08}, nonlinear rate-sensitive yielding \cite{bagnold54,dacruz05,midi2004dense,jop06}, and nonlocality due to the finite size of grains \cite{kamrin2012nonlocal,henann2013predictive,kamrin2015nonlocal,mohan02,aranson02}. Obtaining pragmatic scaling relations from these models or a combination thereof has some inherent difficulties: (1) A quantitative model that attempts to capture the various granular complexities invariably requires more material parameters. These invoke more dimensionless numbers, each of which acts as a constraint on how to make a scaled pair of experiments. (2) Many granular continuum models reference grain properties directly, such as the mean grain diameter. This can lead to scaling relations that require scaled tests to be performed with different grains.  The ability to manufacture such granular systems in practice would be highly nontrivial, and moreover, changing grain size can introduce new physics to the grain-grain interaction, such as the relative importance of charge as grains enter the powder regime. 

Despite the complexities of the constitutive behavior of grains, recent results suggest that the specific problem of calculating resistive forces on generally-shaped, rigid intruders is often well-described with a simple model called the  Resistive Force Theory (RFT) \cite{li13}.  RFT is an empirical model utilizing a set of hypotheses about local drag forces to approximate resistance on general solid surfaces moving in granular soils near the surface.  RFT was initially developed for viscous drag problems \cite{lighthill75}, however it has shown surprising effectiveness in granular media, where it has been used to simulate the dynamics of legged reptiles and robots \cite{li13}, swimming sandfish \cite{goldman14}, and the distribution of lift forces on curved bodies submerged in grains \cite{ding11}.  

In light of its effectiveness in multiple geometries and its dependence on very few model parameters, a natural question to ask is whether granular RFT, when combined with Newton's laws, produces a set of intruder dynamics possessing \edit{scaling behaviors}.  If these could be identified and validated experimentally, they would provide a physical basis to directly relate different granular locomotion problems in the same soil without the need to run any models or perform any discrete simulation. In application, they could be exploited as scaling laws to predict the performance of a ``large'' locomotor in a sand bed --- such as a truck wheel or a tank --- by appropriately down-scaled analysis of a smaller locomotor in the same bed.  Such a capability could be leveraged to produce scaling methods for soil interactions similar to those exploited in aerodynamics and hydrodynamics.  

Herein, we study arbitrarily shaped wheels and derive and experimentally validate a family of geometrically-general scaling laws governing their driving behaviors.  The relations are obtained through analysis of the RFT terradynamical system. As a secondary justification, we show that the same scalings arise by modeling the grains as a simple, rate-independent, frictional continuum.  This raises important questions about the importance of the constitutive complexities of granular flow when studying the specific problem of granular intrusion forces.  It also supports recent results \cite{askari2016intrusion} showing that RFT modeling assumptions emerge strongly in intrusion through a hypothetical frictional-plastic media. A key point about the scaling relations we propose is that they do not require changing the grains themselves in order to relate two experiments with different locomotors. 

The analysis we provide could be applied to locomotors with more moving parts, but here we test the concept on solid wheels for simplicity.  The scaling forms admit variation in the ambient gravity as well, which could have a number of applications in extra-planetary studies; this would admit one to perform a test on earth that predicts how a locomotor functions, say, on Mars.  This ability is validated with discrete element method (DEM) tests, which enable control over gravity.  They show that the scaling relations work over a wide range of locomotive behaviors even as gravity varies between pairs of scaled tests. 

\begin{figure}[t!!!!!!]
\includegraphics[width=0.74\columnwidth,trim={5.8cm 8.6cm 6.4cm 7.6cm},clip]{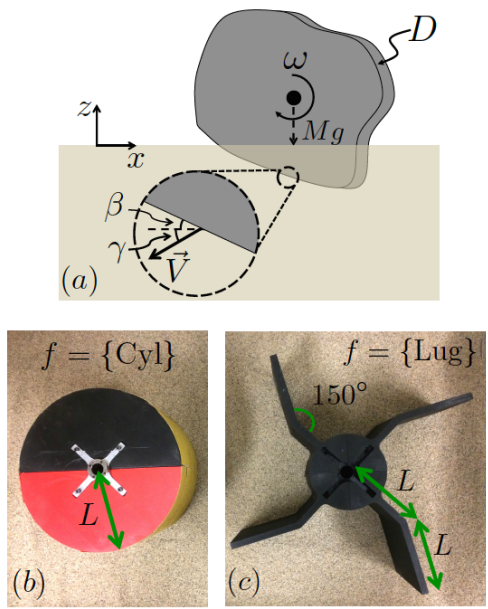}
\caption{\edit{(a) General problem: Driving an arbitrarily shaped wheel of width $D$ and rotational velocity $\omega$, carrying a mass $M$ under gravity $g$. Blow-up of a surface element shown. (b),(c) The two wheel shapes used in our study and their corresponding definitions of $L$.}}
\label{genwheel}
\end{figure}

\section{Resistive Force Theory}
Granular RFT is concerned solely with computing the resistive force on a body moving through a gravitationally loaded bed of grains.  It does not attempt to describe the flow or stress fields in the granular media.  Its basic premise, which is not actually derived physically, though new efforts are being made to provide its foundations \cite{askari2016intrusion}, is that \edit{the stress $(\sigma_x,\sigma_z)$ acting on any surface element of an intruder is determined only by the element's motion direction ($\gamma$), tilt ($\beta$), and depth $|z|$ (see Fig \ref{genwheel}(a)).  Particularly, it is assumed that $\sigma_x=\alpha_x(\beta,\gamma)\cdot\xi \cdot|z|$ and $\sigma_z=\alpha_z(\beta,\gamma)\cdot\xi \cdot|z|$ where $\alpha_x$ and $\alpha_z$ are dimensionless empirical functions with a universal known shape across many materials \cite{li13}, and $\xi$ is a ``grain-structure coefficient.''}  With units of force per volume, $\xi$ is the only parameter needed to execute an RFT calculation.  It depends on the grains, the intruder roughness, and the value of gravity.  By contrast, common engineering terradynamics models like that of Wong and Reece (based on Bekker's work) \cite{wong67} or the NATO Reference Mobility Model \cite{ahlvin92} require a larger number of fit parameters, though they model more than resistive forces.  Unlike Bekker's model \cite{bekker}, no knowledge of the material deformation is required for RFT.

\section{Dimensional Analysis}
The generic wheel has a dimensionless shape, which we denote by a point-set $f$, a constant width $D$ into the plane, a characteristic length $L$ that scales the shape $f$ to give the actual wheel cross-section, and a mass $M$ assumed concentrated on the axle.  The wheel is given a fixed rotational velocity $\omega$, is acted upon by some gravity $g$, and interacts with the sand bed through some grain-structure coefficient $\xi$.  The outputs we are interested in are the power expended as the wheel drives in granular media, $P$, and the wheel's  $x$-translational velocity $V$, although other outputs could be considered.  Both outputs are functions of time, $t$. Before applying dimensional analysis, note that 
\edit{since the wheel surfaces do not vary in the through-thickness direction, when integrating $(\sigma_x,\sigma_z)$ over the wheel surface the dependence of the resistive force on $\xi$ and $D$ is seen only through the product $\xi D$.}  With this and a standard nondimensionalization, the wheel's steady driving limit-cycle is predicted to obey the form:
\begin{equation}
\left[\frac{P}{Mg\sqrt{Lg}},\frac{V}{\sqrt{Lg}}\right]=\boldsymbol{\Psi}\left(\sqrt{\frac{g}{L}} t,f,\frac{g}{L\omega^2},\frac{\xi D L^2}{Mg}\right)
\label{RFTscalND}
\end{equation}
$\boldsymbol{\Psi}$ is a four-input, two-output function as shown. If the gravity, wheel surface texture, and granular media are fixed, $g$ and $\xi$ can be absorbed into the function, giving the reduced form:
\begin{equation}
\left[\frac{P}{M\sqrt{L}},\frac{V}{\sqrt{L}}\right]=\tilde{\boldsymbol{\Psi}}\left(\sqrt{\frac{1}{L}} t,f,\frac{1}{L\omega^2},\frac{D L^2}{M}\right)
\label{RFTscalD}
\end{equation}
The above implies the following family of scaling relations: Consider two experiments with the same $f$, $g$, and $\xi$, but one has inputs $(L,M, D,\omega)$  and the other has inputs $( L',M',D',\omega')=( r L, s M, sr^{-2}D, r^{-1/2}\omega)$ for any positive scalars $r$ and $s$. Then the corresponding driving cycles should obey $\langle P'\rangle=s r^{1/2}\langle P\rangle$ and $\langle V'\rangle=r^{1/2}\langle V\rangle$  where $\langle\cdot\rangle$ denotes a time-average.\\

\section{Connection to Coulomb Plasticity}
Although RFT is empirical, interestingly, Eq \ref{RFTscalD} can also be deduced mechanically by considering wheel motion in a 3D bed of ideal Coulomb material \cite{nedderman}. Here, the grains are treated as a rate-independent frictionally yielding continuum. Such a model could be used to predict the entire sand motion field, but instead consider dimensional analysis implications that can be identified without solving for flow. The inputs to the problem are the wheel descriptors and now three material parameters from the continuum model: the density of the material $\rho$, the material's coefficient of internal friction $\mu$, and the coefficient of sliding friction of the wheel-material interface $\mu_w$.  Hence, wheels driving through this hypothetical continuum must obey the following dimensionless form:
\begin{equation}
\left[\frac{P}{Mg\sqrt{Lg}},\frac{V}{\sqrt{Lg}}\right]=\boldsymbol{\Psi}_\text{Coul}\left(\sqrt{\frac{g}{L}} t,f,\frac{g}{L\omega^2},\frac{D}{L},\frac{\rho L^3}{M},\mu,\mu_w\right)
\label{CoulscalND}
\end{equation}

This can be further reduced, if it is assumed that granular motion under the wheel is approximately invariant in the out-of-plane dimension.  
In this case, if the mass $M$ and width $D$ of the wheel are scaled by some $C_0$, this would be identical to running $C_0$ copies of the wheel side by side.   
The resulting power would be $C_0 P$ and the velocity would remain unchanged.  From Eq \ref{CoulscalND}, this means $\boldsymbol{{\Psi}}_\text{Coul}$ is unchanged under such a transformation, which 
constrains $\boldsymbol{{\Psi}}_\text{Coul}$ to depend on $M$ and $D$ only through the ratio $D/M$, requiring $\boldsymbol{{\Psi}}_\text{Coul}$ to depend on $D/L$ and $\rho L^3/M$ only through their product:
\begin{equation}
\left[\frac{P}{Mg\sqrt{Lg}},\frac{V}{\sqrt{Lg}}\right]=\bar{\boldsymbol{\Psi}}_\text{Coul}\left(\sqrt{\frac{g}{L}} t,f,\frac{g}{L\omega^2},\frac{\rho D L^2}{M},\mu,\mu_w\right)
\label{RFTCoulscalD}
\end{equation}
This form is identical to Eq \ref{RFTscalD} when $g$, $\rho$,  $\mu$, and $\mu_w$ are fixed so that they can be absorbed into the function:
\begin{equation}
\left[\frac{P}{M\sqrt{L}},\frac{V}{\sqrt{L}}\right]=\tilde{\boldsymbol{\Psi}}_\text{Coul}\left(\sqrt{\frac{1}{L}} t,f,\frac{1}{L\omega^2},\frac{D L^2}{M}\right)
\label{RFTCoulscalDabs}
\end{equation}
Therefore, the scalings implied by RFT also arise if one assumes the grains obey Coulomb plasticity. This connection presumes the flow under the wheel is invariant in the out-of-plane dimension, per a wheel with large enough $D$ relative to sinkage.  \\

\section{Experiments}
\begin{figure}[t]
\begin{center}
\includegraphics[width=\columnwidth,trim={0.0cm 0.0cm 0.0cm 0.0cm},clip]{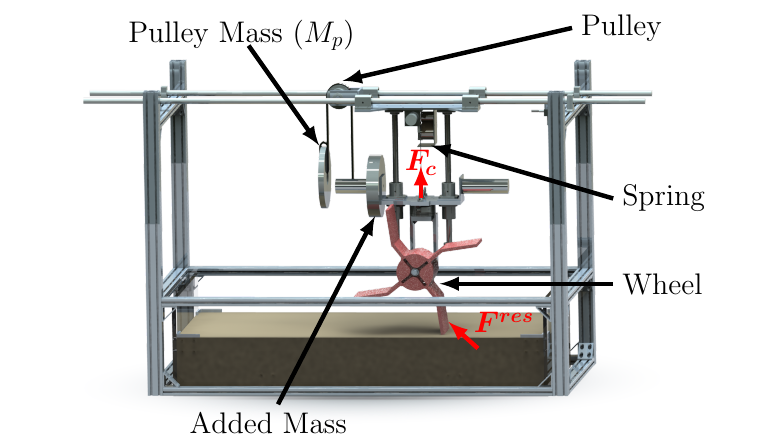}
\end{center}
\caption{Experimental apparatus.  The wheel rotates in the counter-clockwise direction.  A constant force ``spring'' and a pulley are used to alter the wheel's effective gravity.}
\label{ExpSetUp}
\end{figure}

To test the above scalings, an experimental sand bed in the MIT Robotic Mobility Group Lab \cite{senatore2014analysis} was used.  The test apparatus, as shown in Fig \ref{ExpSetUp}, consists of a Lexan bin filled with Quikrete medium sand surrounded by an aluminum frame.  Attached to the aluminum frame are two low friction rods, which are attached to the carriage and allow for horizontal motion.  The carriage is also attached through low-friction vertical rods to the main platform to allow for vertical wheel motion.  This allows the wheel to translate freely, e.g. non-circular wheels ``bobble'' in the vertical direction as they drive through the grains.
The main platform is connected to a motor driven wheel.  The wheels were printed in PLA plastic using a MakerBot 3D printer.  Position, torque, velocity, and rotational velocity sensors are attached to the frame to record the relevant data.

Additionally, we developed a setup (Fig \ref{ExpSetUp}) that allows us to vary the effective gravity on the wheel.  This lets us check Eq \ref{RFTscalND} under different selections of $g$, while keeping $\xi$ held fixed. Lowering gravity on the wheel also prevents overload of the wheel motor as mass increases.   An SDP/SI Neg'ator constant-force ``spring'' ($F_c=66.7$N) is attached between the carriage and the main platform.  A pulley is attached to the carriage platform and a belt around the pulley is attached to the main platform and a mass, $M_p$.  In view of Fig 2,  the gravity change can be seen by applying Newton's second law to the main carriage coupled to the hanging weight.  The result is that the wheel's translational motion always matches that of a free wheel whose mass is $M=M_T+M_p$ and gravity is:
\begin{equation}
g=\frac{M_Tg_e-M_pg_e-F_c}{M_T+M_p}
\label{PulleyMass}
\end{equation}
where $g_e$ is earth gravity and $M_T$ is the total mass of the wheel, motor, main platform, and added mass.  This conversion ensures the wheel's gravitational and inertial masses always agree but does not change the gravity experienced by the grains; the grains are still loaded by earth gravity. Hence, the $\xi$ value remains fixed. Consequently, when we use the conversion technique above to relate two tests, the same effective $g$ must be used for both tests. The upcoming section on Potential Extraplanetary Applications uses discrete element simulations, which allow us to test scalings that arise as the true ambient gravity, acting both on the grains and the wheel, is varied between pairs of tests. 

\begin{figure}[t!!!!!!!!!!!!!!!!!!!!]
(a) 
\includegraphics[width=0.6\columnwidth,trim={2.0cm 9.3cm 2.7cm 8.5cm},clip]{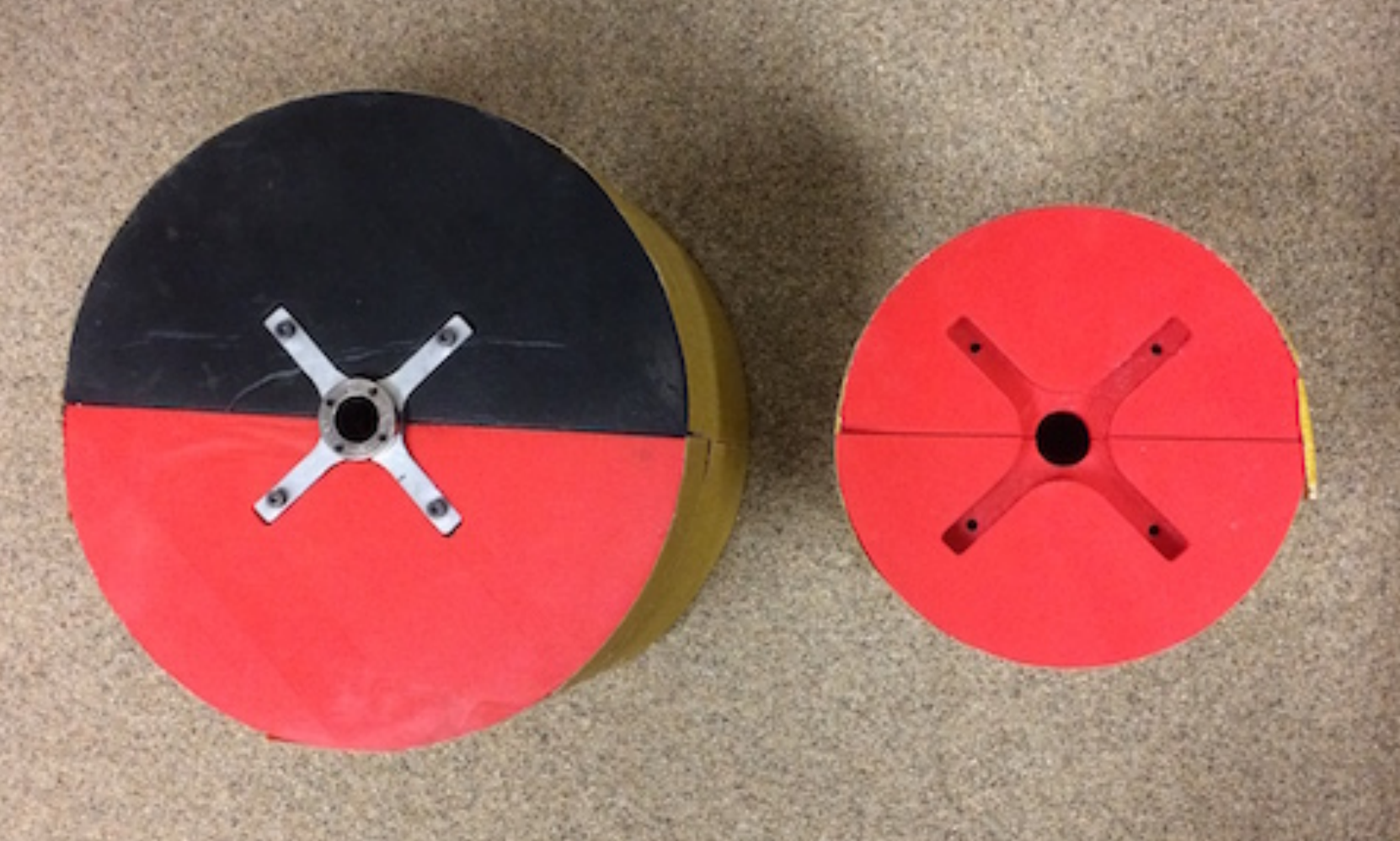} 
\\
\vspace{0.05cm}
(b)  \includegraphics[width=0.6\columnwidth,trim={2.0cm 8cm 2.0cm 8cm},clip]{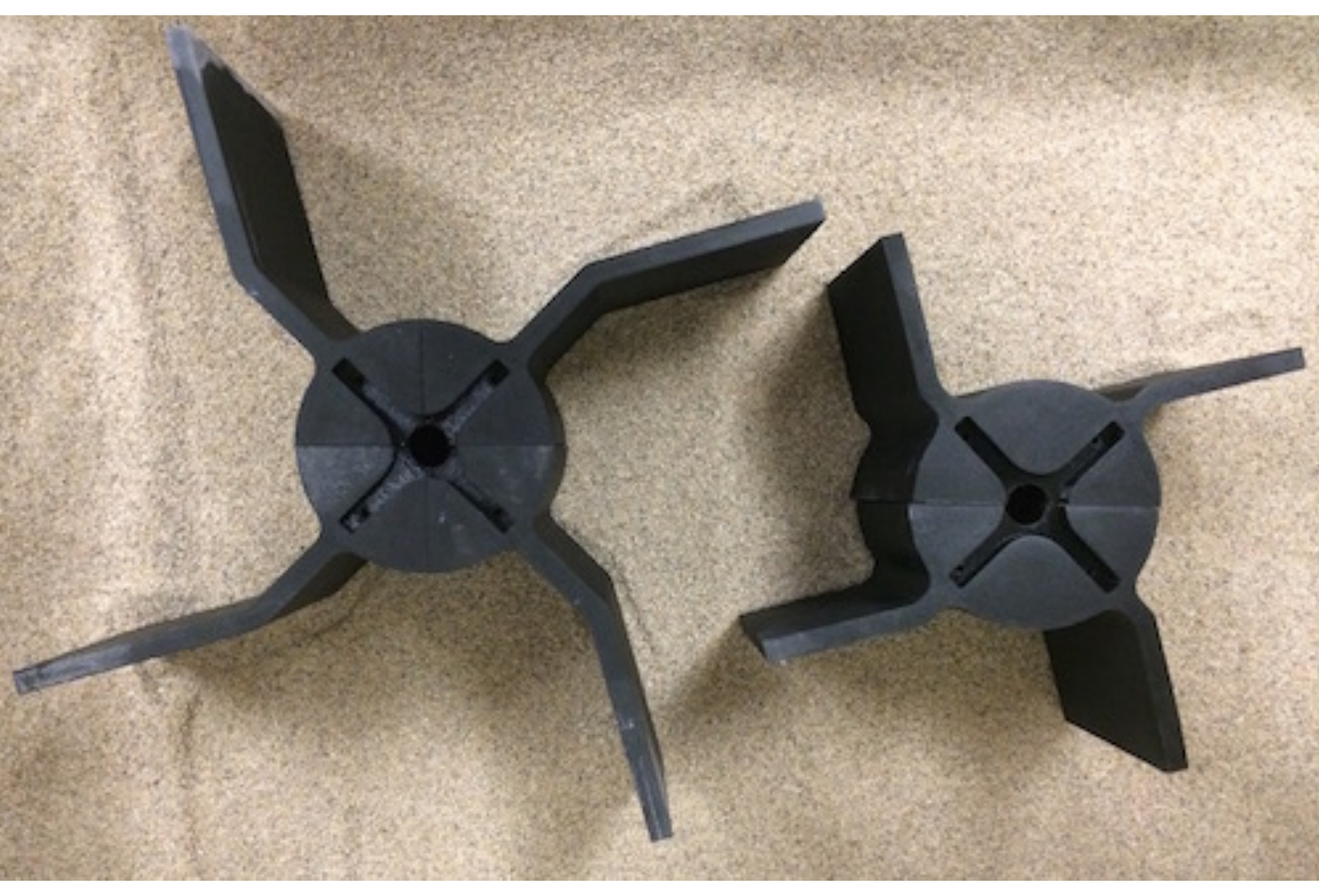} 
\\
\vspace{0.05cm}
(c) \includegraphics[width=0.6\columnwidth,trim={2.0cm 8.5cm 2.0cm 8.5cm},clip]{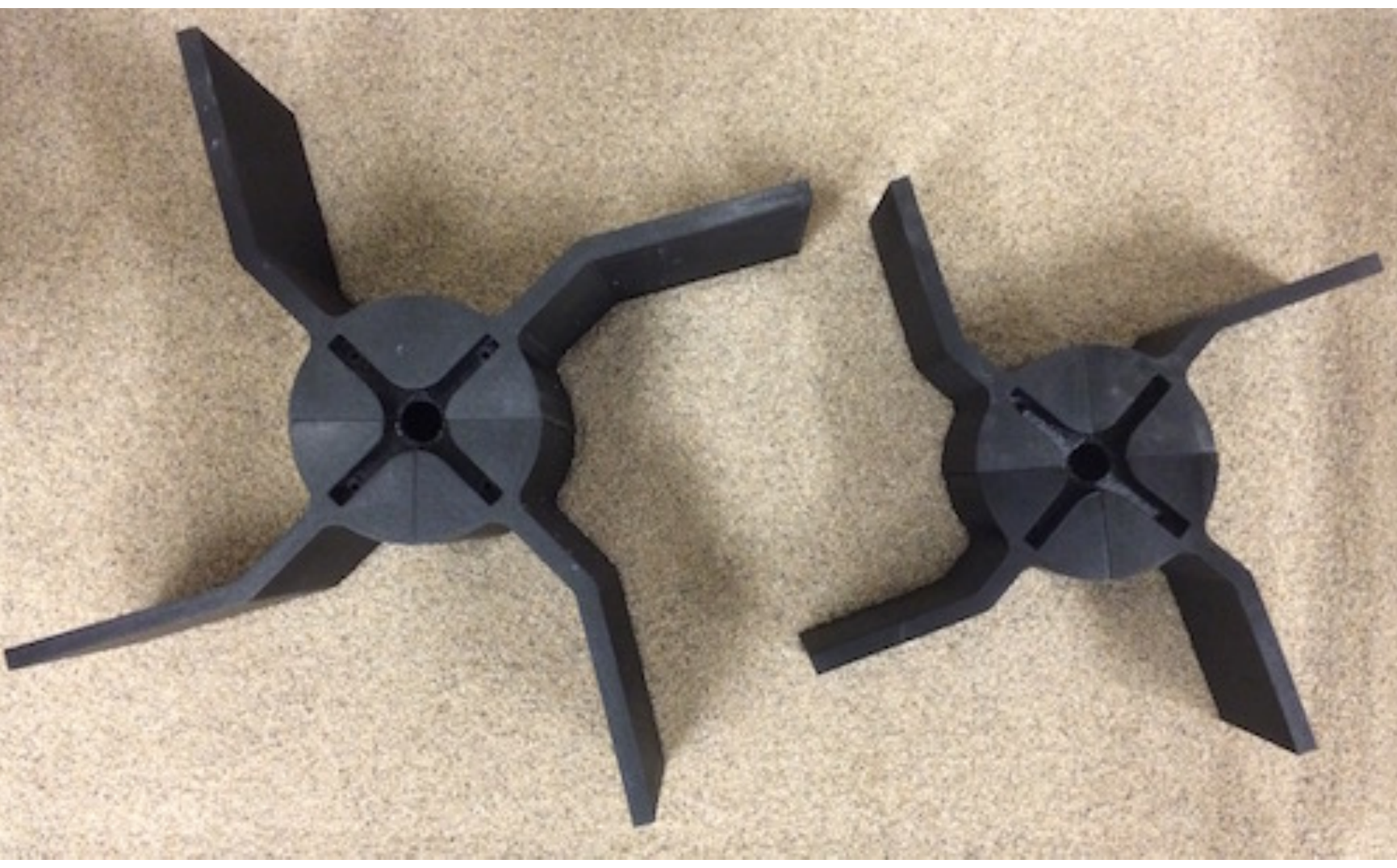}
\caption{The three different wheel pairs tested: Pair A (a), Pair B (b), and Pair C (c).}
\label{WheelPairs}
\end{figure}

The scalings from Equation \ref{RFTscalND} were tested with a set of 288 experiments involving pairs of lug or cylindrical wheels (Fig \ref{WheelPairs}) having different size dimensions, mass loadings, and rotation speeds \footnote{The interior circle on the lug wheels is for mounting and never came into contact with the sand in any tests.}.  \edit{Cylindrical wheels are always} covered in sandpaper to increase interface friction.  Lug wheels have four arms and an elbow half way down their lengths bent 150$^{\circ}$\edit{; previous work suggests an elbow bend improves wheel efficiency \cite{li13}.  The two wheel shapes (cylindrical and lug) were chosen to demonstrate the scaling over two distinct driving motions; cylindrical wheels drive smoothly with limited sinkage, while the lug wheels dig through and remove pockets of sand as they translate (see supplemental movies).} 

\begin{table}
\caption{Experimental tests performed and inputs used.  Each row represents a different wheel pair for which 12 combinations of driving parameters were applied; mass and rotational velocity pairs represented as cartesian products.}
\begin{center}
\begin{tabular}{| >{\centering\arraybackslash} m{0.6cm}| >{\centering\arraybackslash} m{3.8cm}| >{\centering\arraybackslash} m{3.7cm}|}
\hline
 & Common Parameters \: [cm], [cm], [m/s$^2$], [ ] & Driving Parameters \newline [kg] x [deg/s] \\
\hline
Key & $\scriptscriptstyle\begin{bmatrix} L_b\\ L_s \end{bmatrix}$,$\scriptscriptstyle\begin{bmatrix} D_b\\ D_s \end{bmatrix}$,$\scriptscriptstyle\begin{bmatrix} g \end{bmatrix}$,$\displaystyle\left(\{\text{shape}\}\right)$  & \:\: {$\scriptscriptstyle\left(\begin{bmatrix} M_b^1\\ M_s^1 \end{bmatrix}\begin{bmatrix} M_b^2\\ M_s^2 \end{bmatrix}\begin{bmatrix} M_b^3\\ M_s^3 \end{bmatrix}\right)$ x \:\:\:\:\:\:\:\:\:\:\:\:\:\:\:\:\:\:\:\:\:\:\:\:\:\:\:\:\:\:\:\:\:\:\:\:\:\:\:\:\:\:\:\:\:\:\:\:\:\:\:\:\:\:\: $\scriptscriptstyle\left(\begin{bmatrix} \omega_b^1\\ \omega_s^1 \end{bmatrix}\begin{bmatrix} \omega_b^2\\ \omega_s^2 \end{bmatrix}\begin{bmatrix} \omega_b^3\\ \omega_s^3 \end{bmatrix}\begin{bmatrix} \omega_b^4\\ \omega_s^4 \end{bmatrix}\right)$} \:\:\:\:\:\:\:\:\:\:\:\: \\
\hline
Pair A & $\scriptscriptstyle\begin{bmatrix} 12.50\\ 8.08 \end{bmatrix}$,$\scriptscriptstyle\begin{bmatrix} 15\\ 15 \end{bmatrix}$,$\scriptscriptstyle\begin{bmatrix} 3.71 \end{bmatrix}$,$\displaystyle\left(\{\text{Cyl}\}\right)$  & \:\:{$\scriptscriptstyle \left(\begin{bmatrix} 35.9\\ 14.9 \end{bmatrix}\begin{bmatrix} 42.2\\ 17.6 \end{bmatrix}\begin{bmatrix} 45.7\\ 19.0 \end{bmatrix}\right)$ x \:\:\:\:\:\:\:\:\:\:\:\:\:\:\:\:\:\:\:\:\:\:\:\:\:\:\:\:\:\:\:\:\:\:\:\:\:\:\:\:\:\:\:\:\:\:\:\:\:\:\:\:\:\:\:\:\:\:\:\:\:\:\:\:\:\:\:\:\:\:\:\:\:\:\:\:\:\:\:\:\:\:\:\:\:\:\:\:\:\:\:\:\:\:\:\:\:\:\:\:\:\:\:\:\:\:\:\:\:\:\:\:\:\:\:\:\:\:\:\:\:\:\:\:\:\:\:\:\:\:\:\:\:\:\:\:\:\:\:\:\:\:\:\:\:\:\:\: $\scriptscriptstyle\left(\begin{bmatrix} 14.0\\ 17.4 \end{bmatrix}\begin{bmatrix} 17.0\\ 21.2 \end{bmatrix}\begin{bmatrix} 20.0\\ 24.9 \end{bmatrix}\begin{bmatrix} 23.0\\ 28.6 \end{bmatrix}\right)$} \:\:\\
\hline
Pair B & $\scriptscriptstyle\begin{bmatrix} 11.25\\ 7.50 \end{bmatrix}$,$\scriptscriptstyle\begin{bmatrix} 14\\ 14 \end{bmatrix}$,$\scriptscriptstyle\begin{bmatrix} 1.31 \end{bmatrix}$,$\displaystyle\left(\{\text{Lug}\}\right)$  & \:\:{$\scriptscriptstyle\left(\begin{bmatrix} 30.2\\ 13.4 \end{bmatrix}\begin{bmatrix} 34.9\\ 15.5 \end{bmatrix}\begin{bmatrix} 39.7\\ 17.6 \end{bmatrix}\right)$ x \:\:\:\:\:\:\:\:\:\:\:\:\:\:\:\:\:\:\:\:\:\:\:\:\:\:\:\:\:\:\:\:\:\:\:\: $\scriptscriptstyle \left(\begin{bmatrix} 14.0\\ 17.1 \end{bmatrix}\begin{bmatrix} 17.0\\ 20.8 \end{bmatrix}\begin{bmatrix} 20.0\\ 24.5 \end{bmatrix}\begin{bmatrix} 23.0\\ 28.2 \end{bmatrix}\right)$} \:\: \\
\hline
Pair C & $\scriptscriptstyle\begin{bmatrix} 11.25\\ 9.00 \end{bmatrix}$,$\scriptscriptstyle\begin{bmatrix} 14\\ 10 \end{bmatrix}$,$\scriptscriptstyle\begin{bmatrix} 1.31 \end{bmatrix}$,$\displaystyle\left(\{\text{Lug}\}\right)$  & \:\: {$\scriptscriptstyle\left(\begin{bmatrix} 29.3\\ 13.4 \end{bmatrix}\begin{bmatrix} 34.0\\ 15.5 \end{bmatrix}\begin{bmatrix} 38.6\\ 17.6 \end{bmatrix}\right)$ x \:\:\:\:\:\:\:\:\:\:\:\:\:\:\:\:\:\:\:\:\:\:\:\:\:\:\:\:\:\:\:\:\:\:\:\: $\scriptscriptstyle\left(\begin{bmatrix} 14.0 \\ 15.6 \end{bmatrix}\begin{bmatrix} 17.0\\ 19.0 \end{bmatrix}\begin{bmatrix} 20.0\\ 22.4 \end{bmatrix}\begin{bmatrix} 23.0\\ 25.7 \end{bmatrix}\right)$} \:\: \\
\hline
\end{tabular}
\end{center}
\label{ExpInputs}
\end{table}

The different test inputs are shown in Table \ref{ExpInputs}. Four repetitions of the same test are performed each time to obtain useful error-bars. In each test, a pair of wheels of the same shape but different size dimensions were driven through the sand; one wheel is `big' and one is `small' as denoted with subscripts $b$ and $s$ in the table.  Three pairs of wheels were used in total as shown in Fig \ref{WheelPairs}.  In each test, the two masses ($M_b$ and $M_s$) and rotation speeds ($\omega_b$ and $\omega_s$) were chosen to ensure the big and small systems have the same dimensionless inputs to $\boldsymbol{\Psi}$.  The time-averaged non-dimensional power $\langle \tilde{P} \rangle=\langle \frac{P}{Mg\sqrt{Lg}}\rangle$ and velocity $\langle \tilde{V} \rangle=\langle \frac{V}{\sqrt{Lg}}\rangle$ were measured when the wheel motion reached cyclic behavior.  \edit{The power was calculated by multiplying the constant rotation speed by the average torque measured over one cycle.  The velocity was measured directly and averaged over one cycle.  The wheels were started in the same position, such that the torque and velocity measurements could be averaged over corresponding cycles.}  In each case,  the proposed scaling relation is satisfied if $\langle \tilde{P}_b \rangle=\langle \tilde{P}_s \rangle$ and $\langle \tilde{V}_b \rangle=\langle \tilde{V}_s \rangle$. 

\begin{figure}[t!!!!!!!!!!!!!!!!!!!!]
\includegraphics[width=\columnwidth,trim={0.0cm 0.0cm 0.0cm 0.0cm},clip]{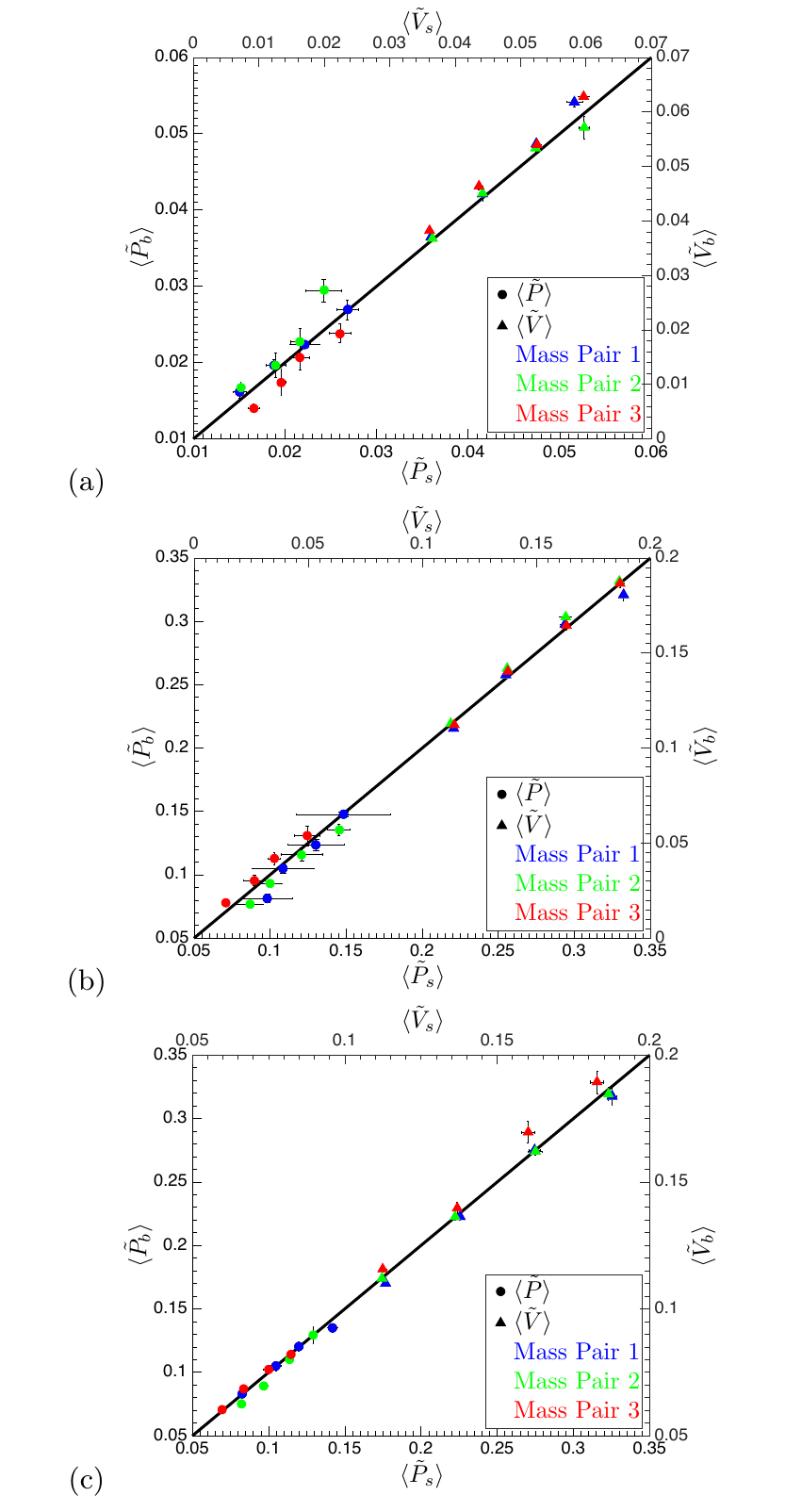}

\caption{Time-average non-dimensional power and velocity values for {Pair A} (a), {Pair B} (b), and {Pair C} (c).  Each color in each plot represents a different mass pair selection and the corresponding four points of each color represent different rotational velocity pairings.  The solid line is the model prediction of $\frac{\langle\tilde{P}_b\rangle}{\langle\tilde{P}_s\rangle}=\frac{\langle\tilde{V}_b\rangle}{\langle\tilde{V}_s\rangle}=1$.
}
\label{alldata}
\end{figure}

The first tests performed were with cylindrical wheels (Pair A).  The non-dimensional powers and velocities arising from each of the 12 pairs of driving parameters are plotted in Fig \ref{alldata}(a).  In general the expected trend is observed at low rotational velocity, with some deviation at the highest rotational rate.
The next series of tests utilized two lug wheels (Pair B).  
Again, the expected scaling relation is observed (Fig \ref{alldata}(b)).
The final series of experiments performed (Pair C) also used four-arm lug wheels, however the widths and lengths of the two wheels were not equal and not proportional.  Hence, each test in this series involves two wheels with different lengths, widths, masses, and rotation speeds --- all differing by different factors --- making this series the most stringent, and arguably most interesting, test of the proposed scaling relation. Strong agreement is observed (Fig \ref{alldata}(c)). The best-fit slopes of all six datasets in Fig \ref{alldata} are all within $3\%$ of the predicted value of 1. 


\edit{In mitigating several laboratory constraints on the size, speed, and mass loading of our wheels, we selected the range of input variables to ensure an ample variation of all inputs rather than a large range of any one variable.   This was done to ensure the dimensionless groups we have identified robustly describe the locomotion process, and that any additional dimensionless groups, which more complex sand models might incur, are not crucial to describe the driving outputs.}

\begin{figure}[h!!!!!!!!!!!!!!!!!!!!!!!!!!!!]
\includegraphics[width=0.55\columnwidth,angle=-90,trim={4.0cm 4.0cm 4.0cm 4.0cm},clip]{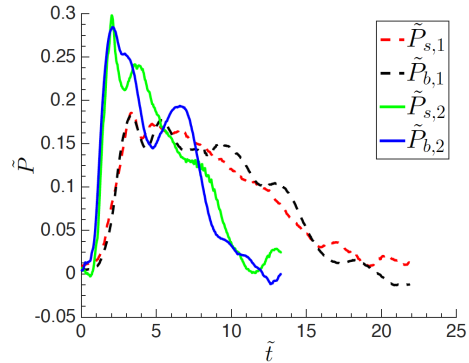}
\caption{Non-dimensional power trajectories for two pairs of lug-wheel experiments over a single cycle \edit{(90$^{\circ}$ of rotation)}.  One pair is shown in solid lines, while the other is shown in dotted lines.  }
\label{SlapScalePowUp}
\end{figure}

From Eq 2, in addition to the time-average behaviors, the actual time-dependence of $\tilde P$ and $\tilde V$ should relate when time is accordingly scaled.  Figure \ref{SlapScalePowUp} shows the non-dimensional power plotted against dimensionless time, $\tilde t = \sqrt{\frac{g}{L}}t$, over the course of a single cycle from two pairs of experiments from wheel Pair C.  The predicted \edit{scaling} agreement between trajectories is observed.  The result is non-trivial; the dimensional powers differ by a factor of 2.45 for each pair. 

\section{Potential Extraplanetary Applications}
It would be advantageous to be able to scale to two experiments in the same sand but with different ambient gravities; e.g. to use an earthbound experiment to predict an extraplanetary experiment. To derive the proper scalings, we can use Eq \ref{RFTCoulscalD}, for which gravity dependence of the granular response is natively accounted.  Interestingly, we can show that the same scalings arise assuming RFT (Eq \ref{RFTscalND}) once the dependence of $\xi$ on gravity is deduced.  
To deduce a form for $\xi$, we assume the grains behave as an ideal Coulomb material, and then dimensional analysis implies the formula for $\xi$ obeys
\begin{equation}
\xi=\rho g \hat\xi\left(\mu,\mu_w\right).
\label{xig}
\end{equation}


By substituting Eq \ref{xig} into Eq \ref{RFTscalND}, we obtain
\begin{equation}
\left[\frac{P}{Mg\sqrt{Lg}},\frac{V}{\sqrt{Lg}}\right]=\boldsymbol{\Psi}\left(\sqrt{\frac{g}{L}} t,f,\frac{g}{L\omega^2},\frac{\rho \hat\xi\left(\mu,\mu_w\right) D L^2}{M}\right).
\label{RFTgrav}
\end{equation}
Note that this relation is equivalent to Eq \ref{RFTCoulscalD} as long as material parameters for the grains are absorbed into the function.  So, even when gravity varies between tests, both approaches again yield the same scaling relations, which can be expressed as follows: Consider two experiments with common wheel shape $f$ and common grains, where one test is described by the inputs $(g, L, M, D,\omega)$  and the other by the inputs $(g', L',M',D',\omega')=(qg, r L, s M, sr^{-2}D, q^{1/2}r^{-1/2}\omega)$ for any positive scalars $q$, $r$, and $s$. Then the steady driving cycles of the corresponding outputs should obey $\langle P'\rangle=q^{3/2} r^{1/2}s \langle P\rangle$ and $\langle V'\rangle=q^{1/2}r^{1/2}\langle V\rangle$.


\begin{figure}
\begin{center}
(a)\ \ \ \ \ \ \includegraphics[width=.5\columnwidth,trim={0.0cm 0.0cm 0.0cm 0.0cm},clip]{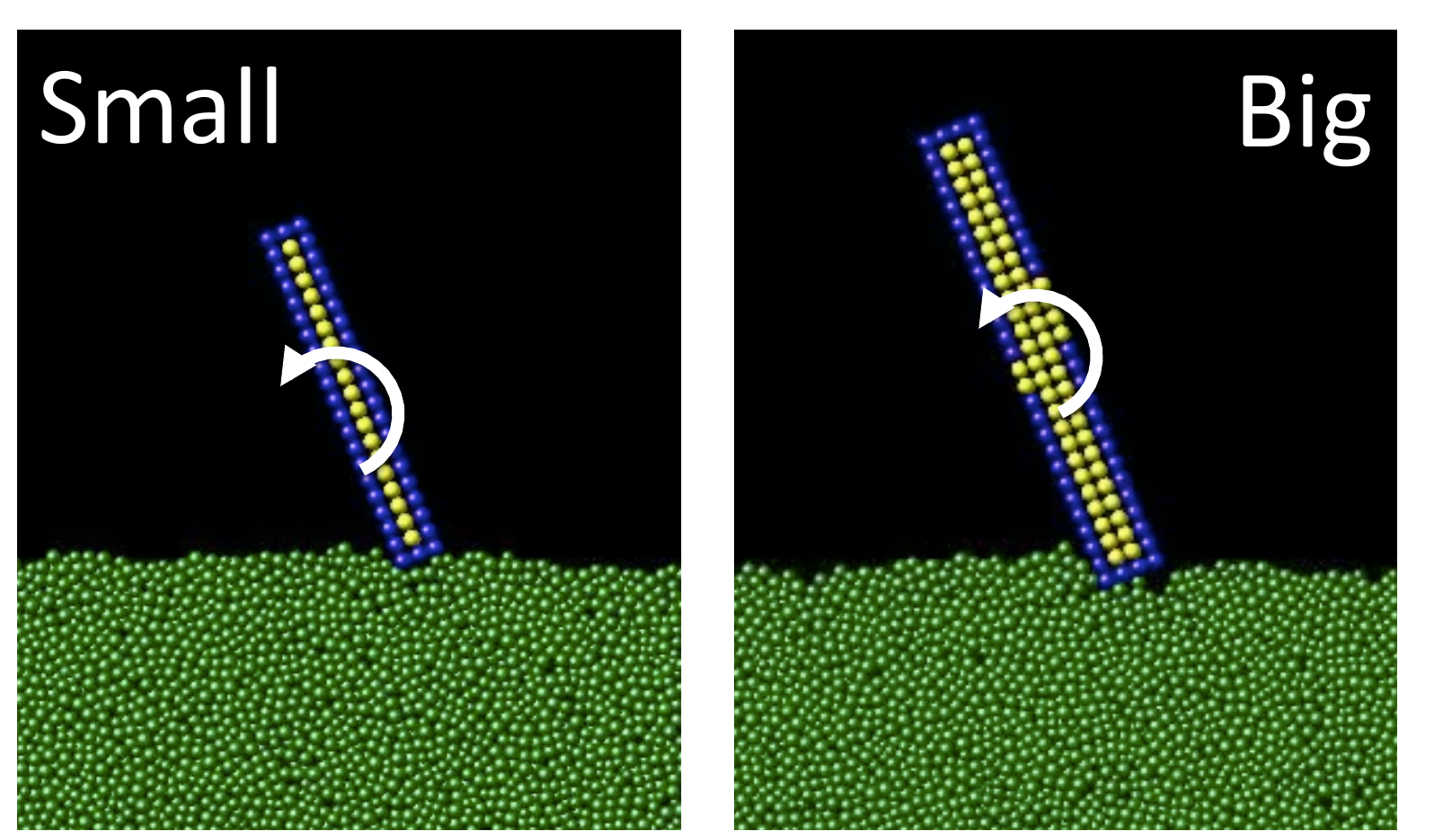}
\\
(b)\includegraphics[width=.6\columnwidth,trim={0.0cm 0.0cm 0.0cm 0.0cm},clip]{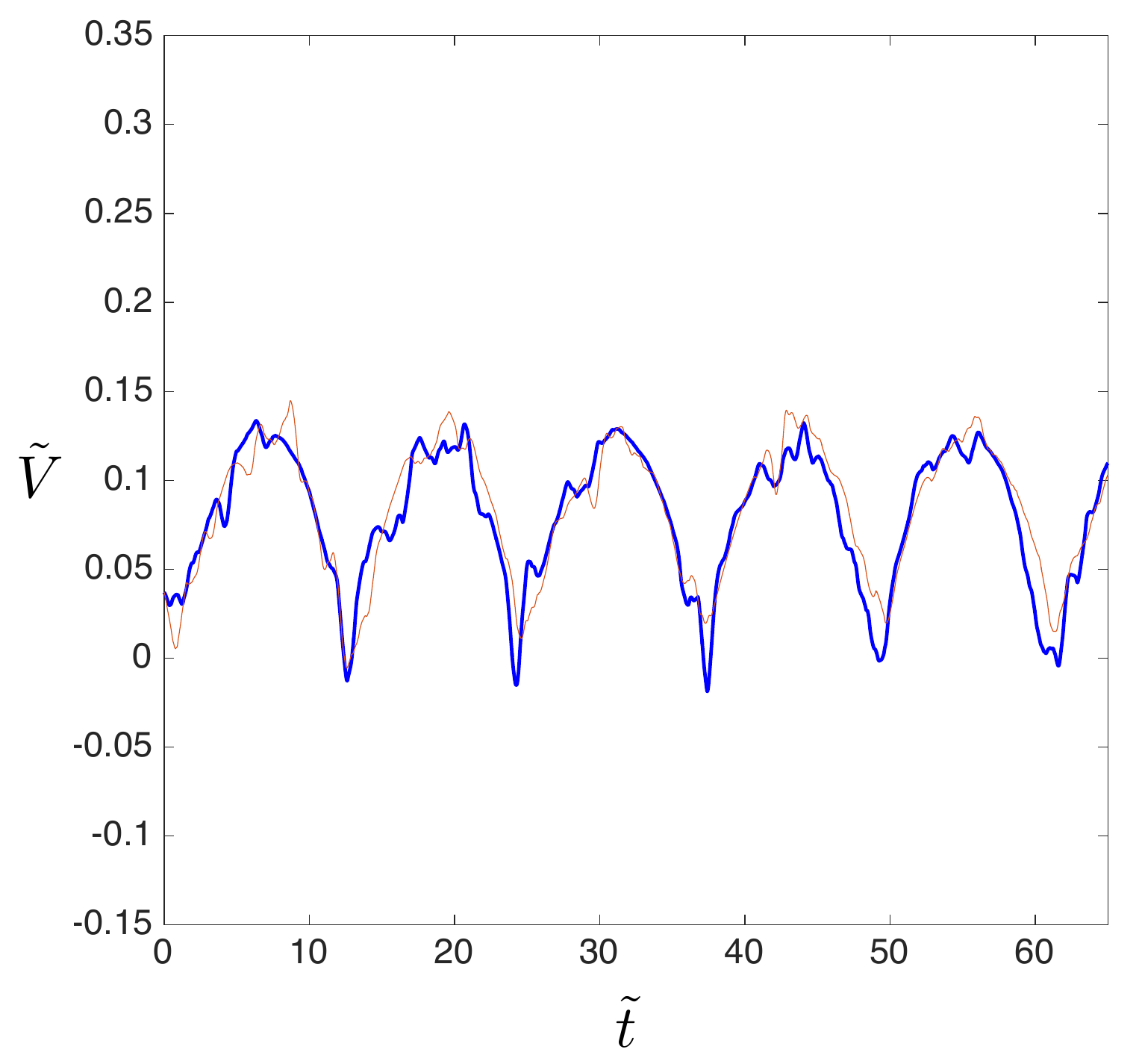}
\\
(c)\includegraphics[width=.6\columnwidth,trim={0.0cm 0.0cm 0.0cm 0.0cm},clip]{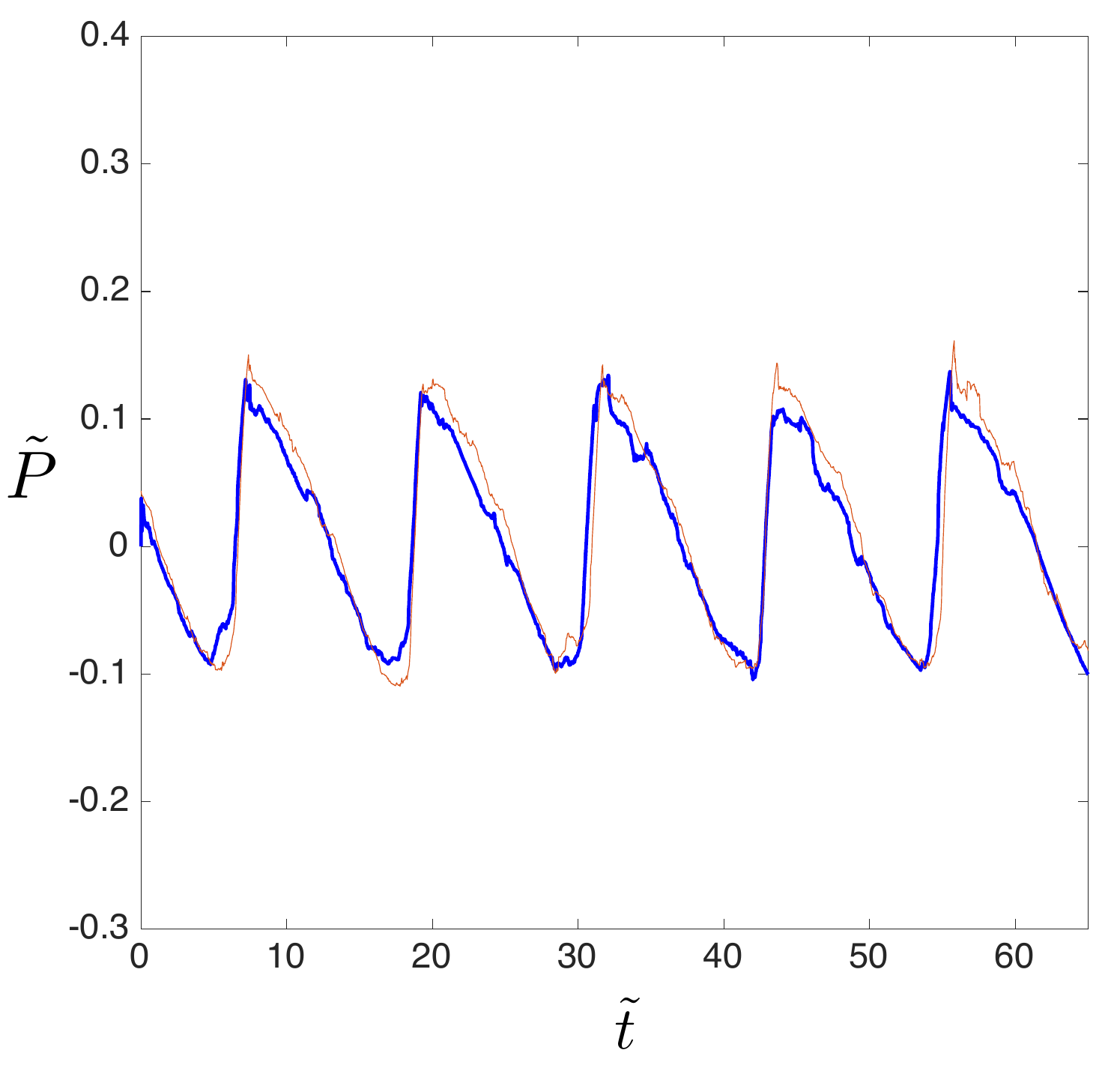}
\end{center}
\caption{\textbf{Walking} mode ($\tilde{\rho}=44.4$, $\tilde{g}=14.8$).  Small test inputs: $g_s=9.8\frac{\text{m}}{\text{s}^2}$, $L_s= .0168\text{m}$, $M_s=12.7\times10^{-6}\text{kg}$, $\omega_s=6.28 \frac{\text{rad}}{\text{s}}$.  Big test inputs: $g_b=39.2\frac{\text{m}}{\text{s}^2}$, $L_b= .0224\text{m}$, $M_b=22.4\times10^{-6}\text{kg}$, $\omega_b=10.9 \frac{\text{rad}}{\text{s}}$. Both tests use the same grains.  (a) Snapshot of the small and big systems at a common value of $\tilde{t}$. (b)-(c)  Time-dependence of power and velocity in dimensionless form during locomotion process. Blue lines are for the big test and red lines for the small.}\label{walk}
\end{figure}

\edit{To see if indeed the above form can be used to relate to tests in two different gravities, we perform a sequence of simulations using Discrete Element Method (DEM). DEM tests allow for the easy manipulation of input parameters, including gravitational acceleration $g$, over a broad scope of values not limited by experimental hardware. We use LAMMPS \cite{plimpton1995fast} for the simulations. Throughout, the grains have the following characteristics: density $\rho_s = 2500$ $\frac{\text{kg}}{\text{m}^3}$, normal and tangential stiffnesses $k_N = 16411 \frac{\text{N}}{\text{m}}$ and $k_t = 4689 \frac{\text{N}}{\text{m}}$, normal damping coefficients $\gamma_N = 184835 \text{s}^{-1}$ and friction coefficient $\mu = 0.4$. The simulations are effectively 2D; the bed is modeled as a planar packing of polydisperse spheres with a linearly distributed radius with minimum  $R_{\text{min}} = 0.254$mm and maximum  $R_{\text{max}} = 0.381$mm. These choices for the particle material have been used successfully in past studies \cite{koval2009annular} and resemble the parameters for glass beads. The wheels are made of a coplanar collection of $0.8$mm diameter particles constrained to move in the plane as a rigid body ($D=0.8$mm).


Per Eq \ref{RFTCoulscalD}, when material is fixed, the key inputs are  $\tilde{\rho}\equiv\frac{\rho D L^2}{M}$, $\tilde{g}  \equiv \frac{g}{L\omega^2}$, and wheel shape $f$.  All DEM tests performed use $f$ corresponding to a fully rough bar rotating about its center, with $L$ the length of the bar.  The scaling law is checked by running pairs of simulations with common values of $\tilde{\rho}$ and $\tilde{g}$. In each pair, the dimensional inputs for the `big' test are given from the `small' by $(g_b, L_b, M_b, \omega_b)=(4g_s,\frac{4}{3}L_s,\frac{16}{9}M_s,\sqrt{3}\omega_s)$. Choosing the factors this way guarantees each of the pair has the same dimensionless inputs according to Eq \ref{RFTCoulscalD}. Since all four inputs differ in each test pair, this offers a stringent test of the scaling relation. We emphasize that the big wheel test has $4\times$ the ambient gravity of the small in each test pair, and the same grains are used among all tests.

The first pairing has small wheel inputs as follows: gravity is $g_s=9.8\frac{\text{m}}{\text{s}^2}$, wheel length is $L_s= .0168\text{m}$, wheel mass is $M_s=12.7\times10^{-6}\text{kg}$, and rotation speed is $\omega_s=6.28 \frac{\text{rad}}{\text{s}})$. The big wheel inputs are $4g_s, \frac{4}{3}L_s,\frac{16}{9}M_s,\sqrt{3}\omega_s$ as noted above. These tests both have $\tilde{\rho}=44.4$, $\tilde{g}=14.8$, and produce a `walking' motion where the wheels are so light compared to the grains they barely deform the granular surface (see supplemental videos).  Figure \ref{walk} displays the results. The scaling law would predict the two tests should have the same $\tilde{V}=V/\sqrt{Lg}$ vs $\tilde{t}$ and $\tilde{P}=P/Mg\sqrt{Lg}$ vs $\tilde{t}$ curves because they share all the same dimensionless inputs to Eq \ref{RFTCoulscalD}. The strong overlap of the $\tilde{V}=V/\sqrt{Lg}$ vs $\tilde{t}=t\sqrt{g/L}$ curves verifies the scaling relation in its full time-dependent form. The snapshot in Fig \ref{walk}(a) shows the two wheels at a common value of $\tilde{t}$. The similarity in the wheel positioning and granular free surface between properly scaled tests at the same $\tilde{t}$ is clear.

The second pairing increases both wheel masses ten-fold, giving $\tilde{\rho}=4.44$, $\tilde{g}=14.8$, which produces a `trudging' motion where the wheels are so heavy they submerge nearly to the central axle while driving through the grains (see supplemental videos).  Results of these tests are presented in Fig \ref{trudge}, which confirms the scaling law's prediction that these two tests should have the same dimensionless power behavior and dimensionless velocity behavior.  Moreover, the snapshot in Fig \ref{trudge}(a) shows that the divit in the free surface created by each wheel looks geometrically similar to the other, which supports the notion that these two problems obey a proper scaling.

\begin{figure}
\begin{center}
(a)\ \ \ \ \ \ \includegraphics[width=.5\columnwidth,trim={0.0cm 0.0cm 0.0cm 0.0cm},clip]{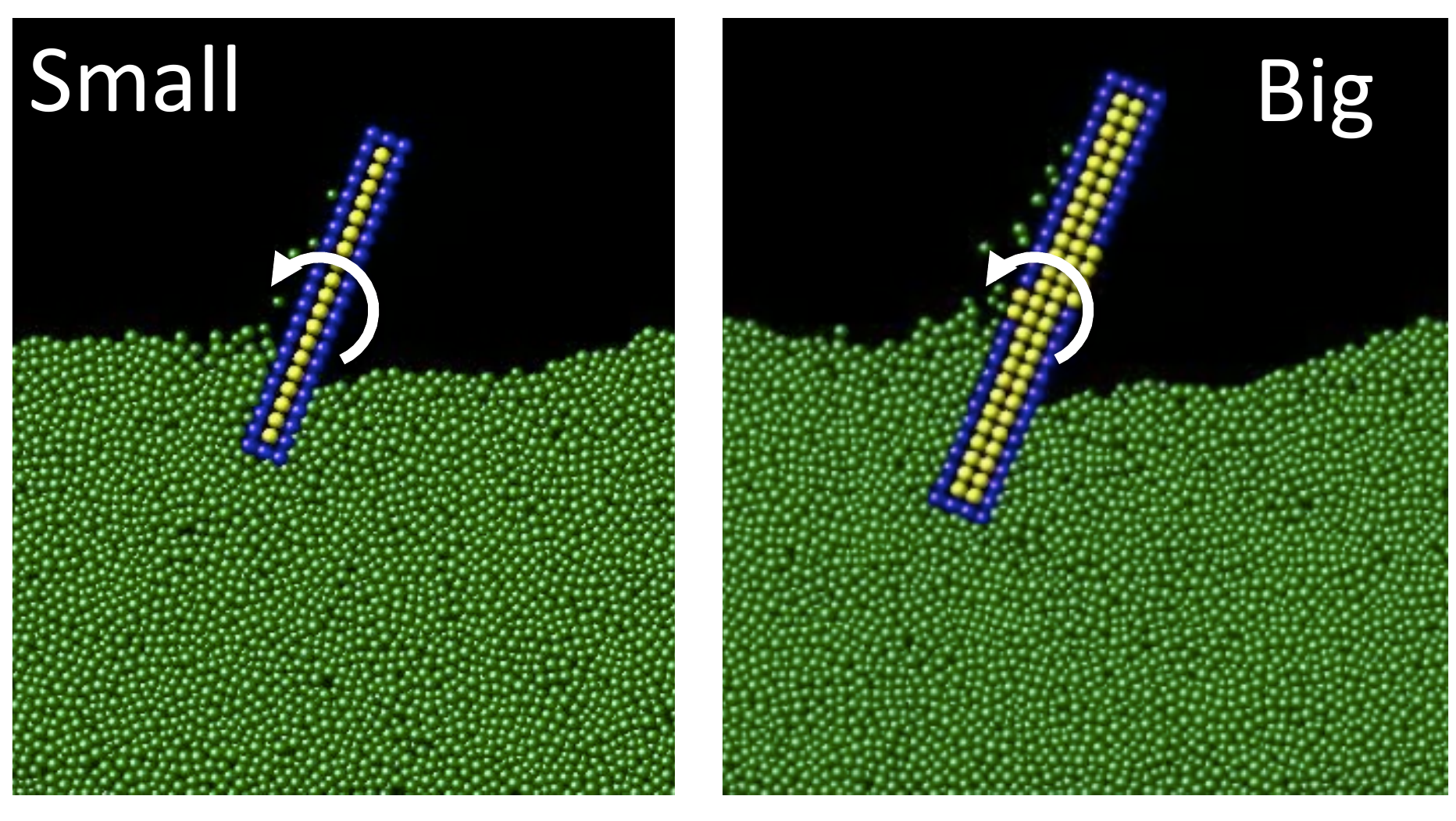}
\\
(b)\includegraphics[width=.6\columnwidth,trim={0.0cm 0.0cm 0.0cm 0.0cm},clip]{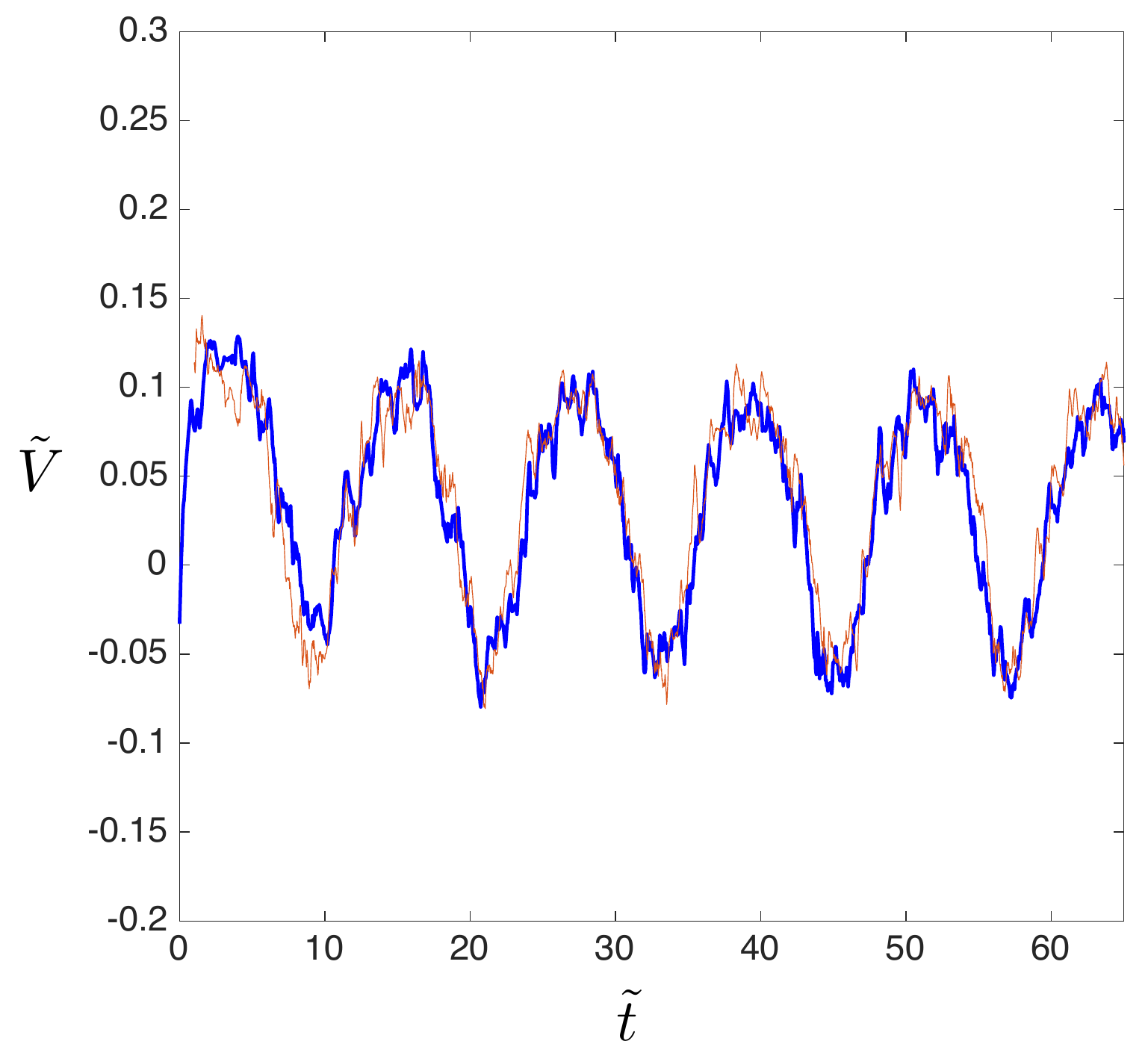}
\\
(c)\includegraphics[width=.6\columnwidth,trim={0.0cm 0.0cm 0.0cm 0.0cm},clip]{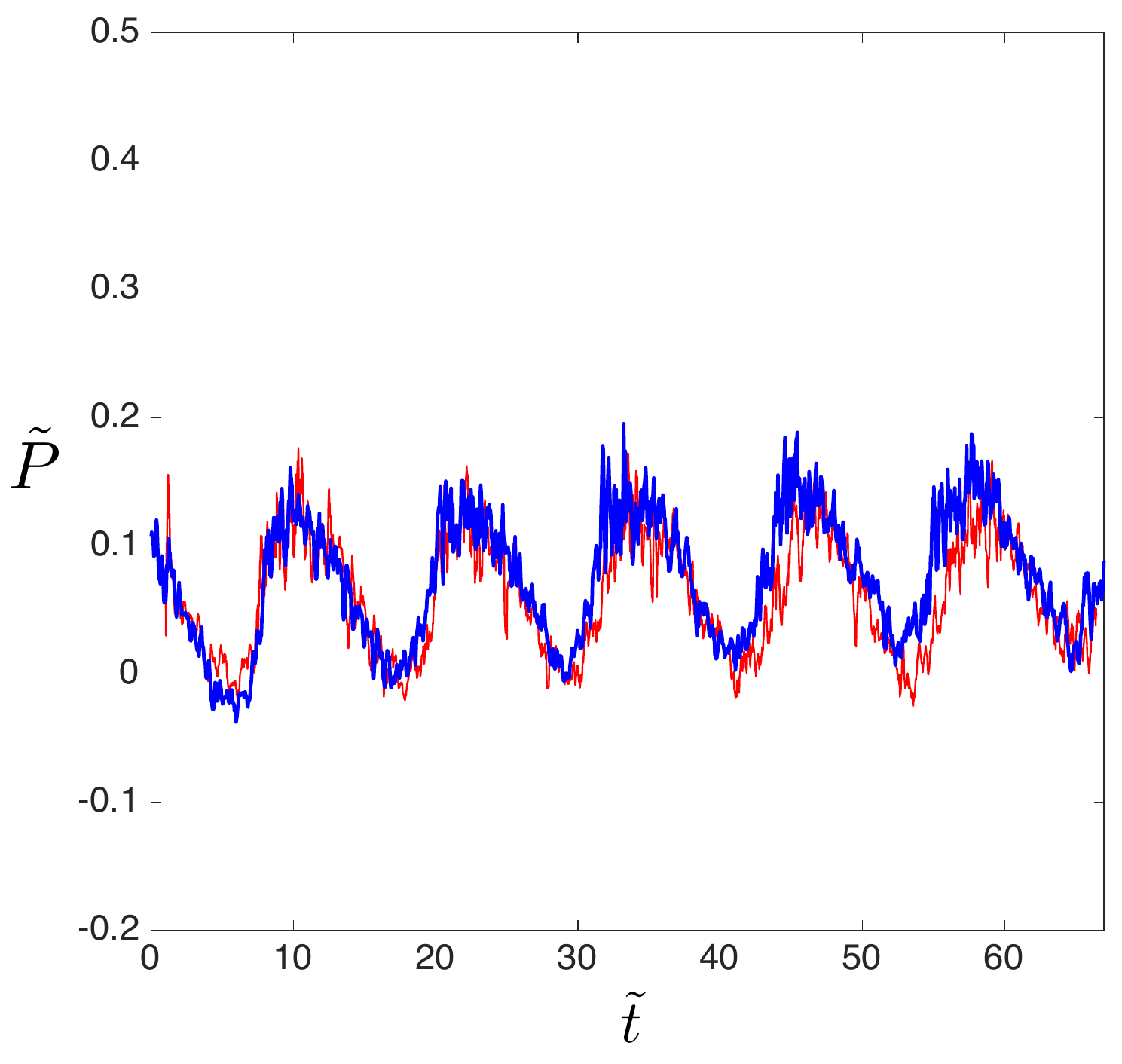}
\end{center}
\caption{\textbf{Trudging} mode ($\tilde{\rho}=4.44$, $\tilde{g}=14.8$). Small test inputs: $g_s=9.8\frac{\text{m}}{\text{s}^2}$, $L_s= .0168\text{m}$, $M_s=127\times10^{-6}\text{kg}$, $\omega_s=6.28 \frac{\text{rad}}{\text{s}}$.  Big test inputs: $g_b=39.2\frac{\text{m}}{\text{s}^2}$, $L_b= .0224\text{m}$, $M_b=224\times10^{-6}\text{kg}$, $\omega_b=10.9 \frac{\text{rad}}{\text{s}}$. Both tests use the same grains. (a) Snapshot of the small and big systems at a common value of $\tilde{t}$. (b)-(c)  Blue lines are for the big test and red lines for the small.}\label{trudge}
\end{figure}

The third simulation pair uses an intermediate pair of masses ($M_s=42.2\times10^{-6}\text{kg}$) but increases the rotation speeds ten-fold, giving $\tilde{\rho}=13.4$, $\tilde{g}=0.148$.  These inputs cause a `skipping' motion where the wheels lift off the bed entirely each cycle and eject grains as they propel forward (see supplemental videos).  Figure \ref{skip} displays the results for the pair of tests.  There is more noise in the data due to the rapid impulses, but the data still indicates a strong agreement between the big and small tests.  Unlike the prior two cases, the skipping case takes a significant time to accelerate to the steady driving motion; acceleration is evident from $\tilde{t}=0$ to $\sim 15$ in Fig \ref{skip}(b).  The snapshot shows that at a common value of $\tilde{t}$ various observations about the two systems look very similar; the free surface shape and ejected grain plume for example look very similar as one might expect for two problems that are properly scaled versions of each other.

\begin{figure}
\begin{center}
(a)\ \ \ \ \ \ \includegraphics[width=.7\columnwidth,trim={0.0cm 0.0cm 0.0cm 0.0cm},clip]{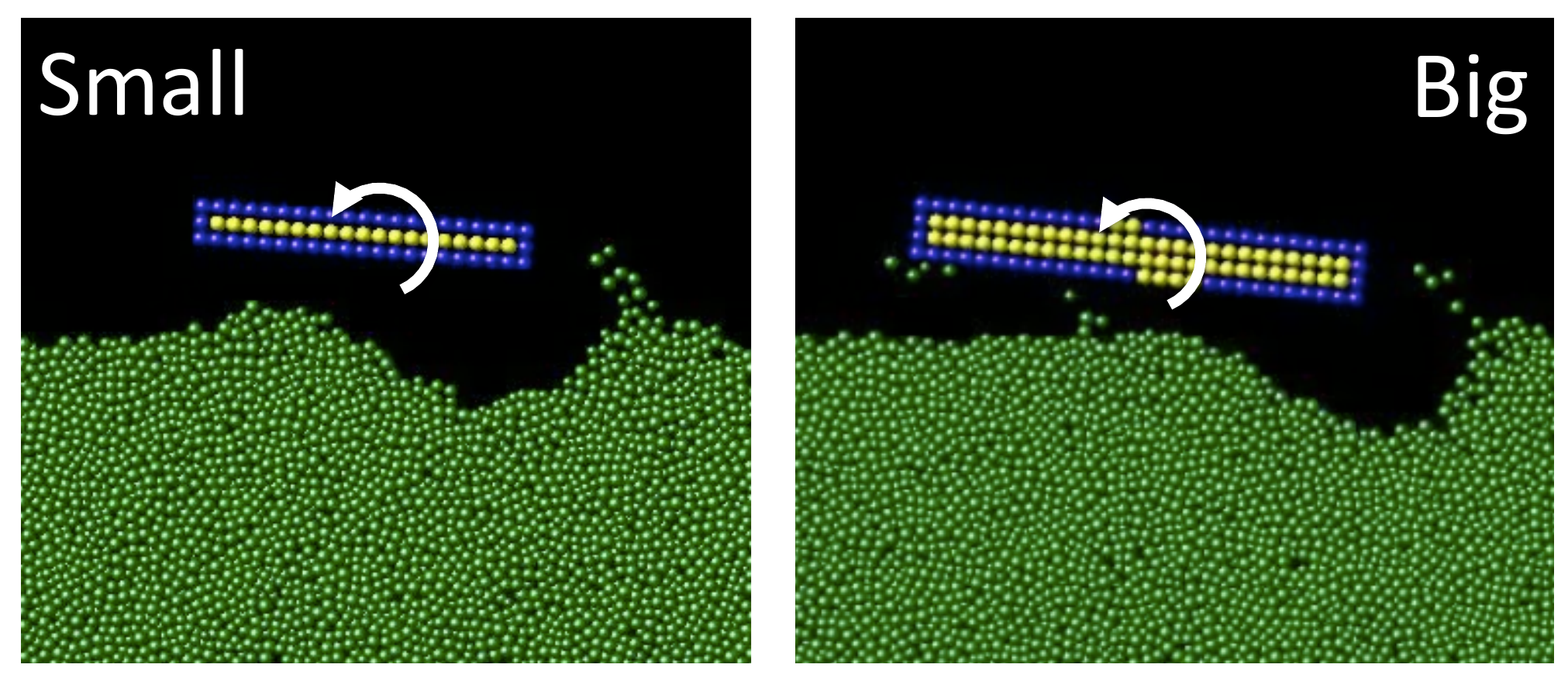}
\\
(b)\includegraphics[width=.6\columnwidth,trim={0.0cm 0.0cm 0.0cm 0.0cm},clip]{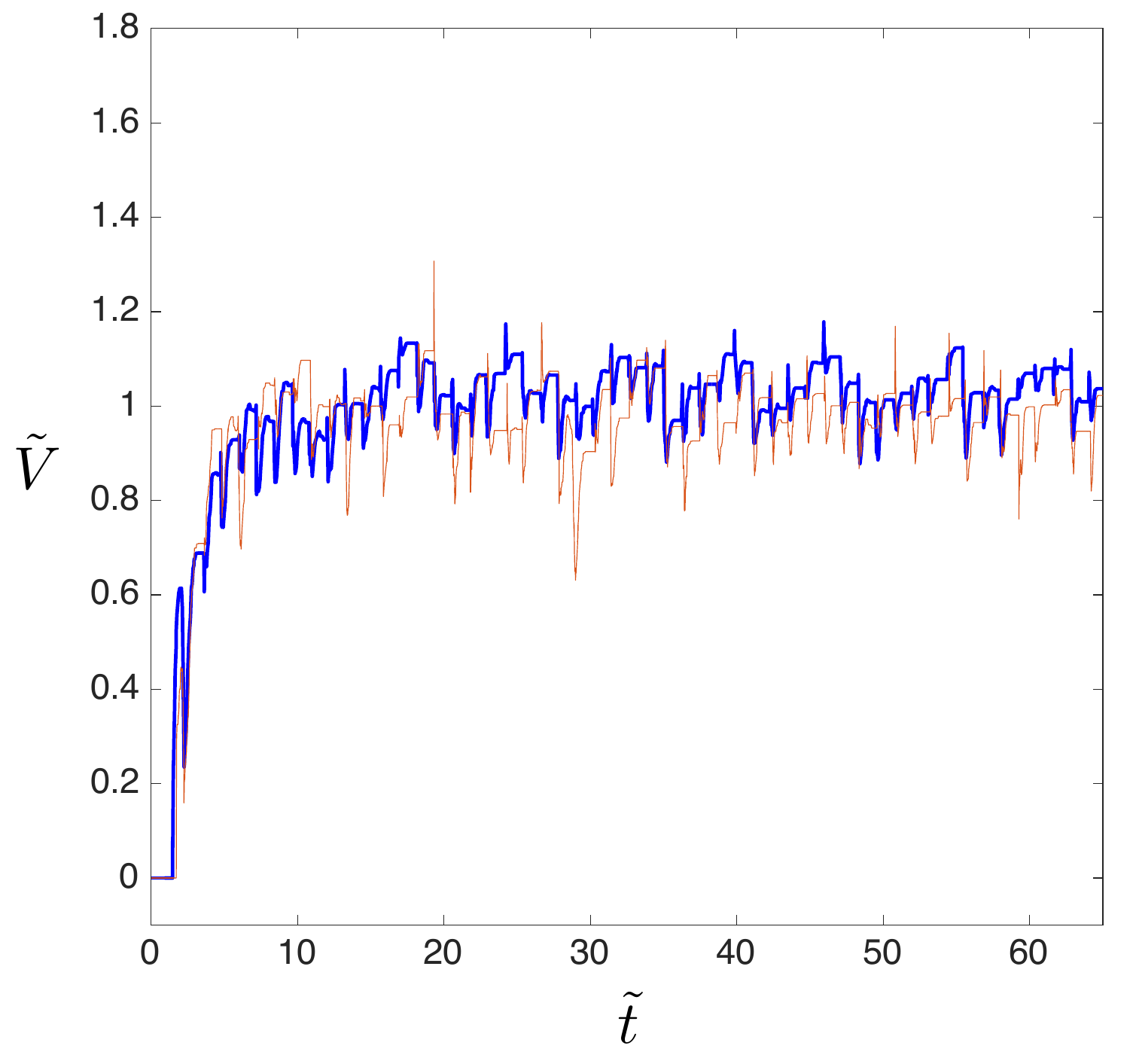}
\\
(c)\includegraphics[width=.6\columnwidth,trim={0.0cm 0.0cm 0.0cm 0.0cm},clip]{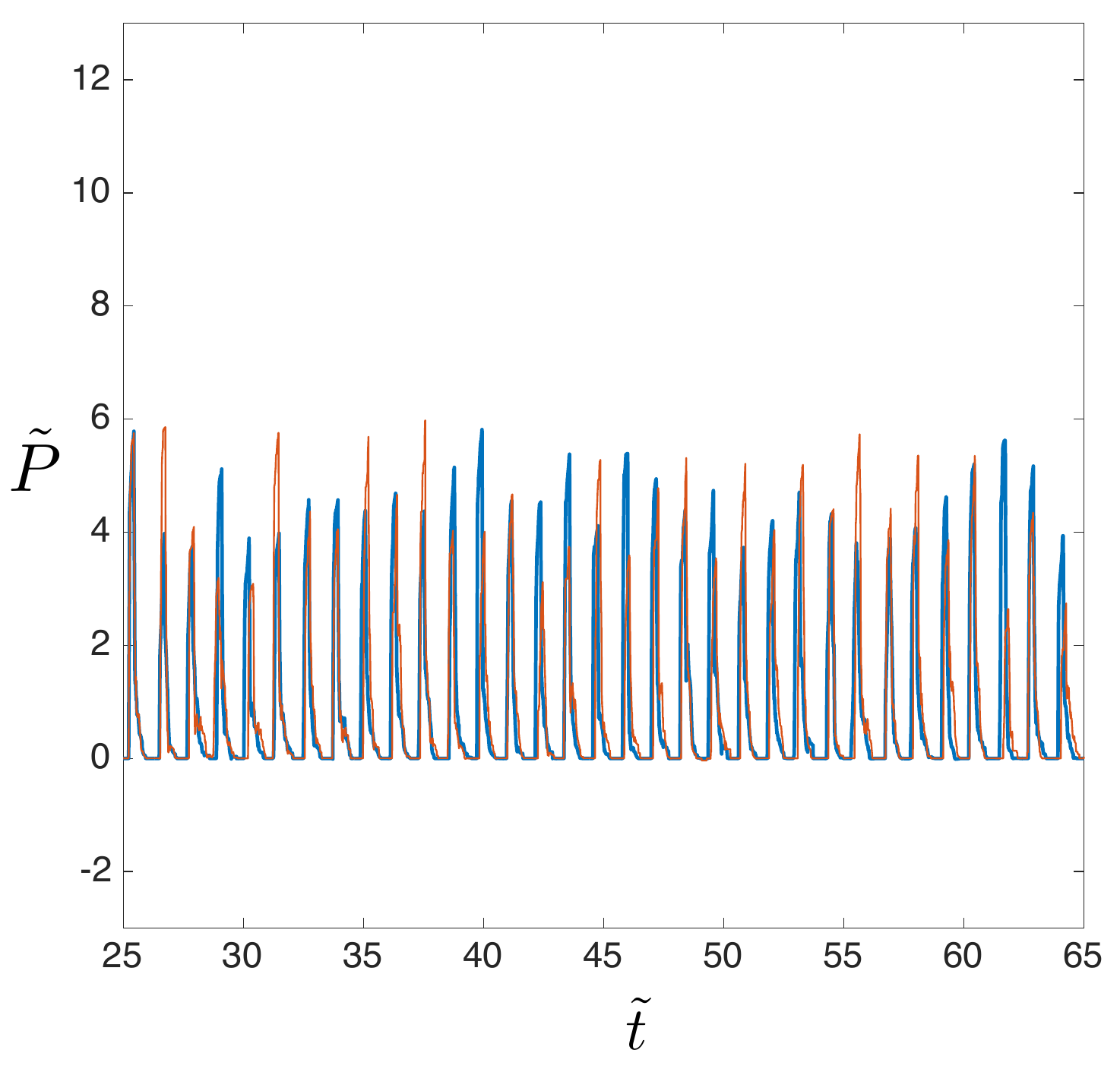}
\end{center}
\caption{\textbf{Skipping} mode ($\tilde{\rho}=13.4$, $\tilde{g}=0.148$). Small test inputs: $g_s=9.8\frac{\text{m}}{\text{s}^2}$, $L_s= .0168\text{m}$, $M_s=42.2\times10^{-6}\text{kg}$, $\omega_s=62.8 \frac{\text{rad}}{\text{s}}$.  Big test inputs: $g_b=39.2\frac{\text{m}}{\text{s}^2}$, $L_b= .0224\text{m}$, $M_b=75.0\times10^{-6}\text{kg}$, $\omega_b=109 \frac{\text{rad}}{\text{s}}$. Both tests use the same grains.  (a) Snapshot of the small and big systems at a common value of $\tilde{t}$. (b)-(c)  Blue lines are for the big test and red lines for the small. }\label{skip}
\end{figure}

The tests we have conducted cover a fairly wide range of qualitative behaviors of the locomotor, and it is reassuring that even as the mode of locomotion changes, the scaling relation still appears to hold, as is evident in Figs \ref{walk}-\ref{skip}.
}



\section{Discussion}
We have proposed and validated a set of scaling laws in granular locomotion for the case of rigid, arbitrarily shaped wheels, which was initially obtained by analyzing RFT system dynamics.  The scaling analysis has been reconciled with Coulomb plasticity and an extension has been \edit{shown} to relate locomotion processes in different ambient gravity.  Our study has shown the scaling relations to work over three different wheel shapes (cylinders, lugs, and bars), and two different choices for the granular material (sand in the experiments, and polydisperse spheres in the DEM tests).

In the future, we would like to extend this scaling to include a wheel's ability to pull a load. This can be described as a constant drawbar force $F_d$, which acts in the backward (negative) horizontal direction.  The drawbar force is useful as an indicator of both tractive performance and ability to scale inclines. With this new consideration, we can extend the relationship found in Eq \ref{RFTgrav} to include $F_d$ by adding an additional non-dimensional group:
\begin{multline}
\left[\frac{P}{Mg\sqrt{Lg}},\frac{V}{\sqrt{Lg}}\right]=\\
\boldsymbol{\Psi}\left(\sqrt{\frac{g}{L}}  t,f,\frac{g}{L\omega^2},\frac{\rho \hat\xi\left(\mu,\mu_w\right) D L^2}{M},\frac{F_d}{Mg}\right)
\label{RFTCoulscalFD}
\end{multline}
Using this new form, we can expound the scaling.  Suppose two experiments with common $f$, $\rho$, $\mu$, and $\mu_w$, but one is described by the inputs $(g, L, M, D,\omega, F_d)$  and the other by the inputs $(g', L',M',D',\omega', F_d')=(qg, r L, s M, sr^{-2}D, q^{1/2}r^{-1/2}\omega, sq F_d)$ for any positive scalars $q$, $r$, and $s$. The steady driving cycles of the corresponding outputs should then obey $\langle P'\rangle=q^{3/2} r^{1/2}s \langle P\rangle$ and $\langle V'\rangle=q^{1/2}r^{1/2}\langle V\rangle$.  It would be interesting future work to experimentally validate this proposed scaling.

In general, there are two possible explanations for why scaling laws arising from rate-independent Coulomb plasticity (or RFT) work as well as they do, even though granular material is known to obey more complicated constitutive behaviors (as described in the Introduction).  First, it is possible that the complexities of granular rheology influence granular kinematics moreso than boundary stresses. Findings in \cite{henann2013predictive} showed that in split-bottom Couette flow, even though the flow features are influenced noticeably by nonlocality due to grain size, the stresses transmitted to the solid boundaries are largely unaffected by nonlocal effects.  This suggests that even if the flow fields predicted by Coulomb plasticity are less accurate than those of more complete models, the intruder forces it predicts may be quite similar.  If so, the intruder dynamics would still be accurately modeled despite using a simplified granular continuum model.  

The second explanation is that locomotion in granular beds, even over a relatively large range inputs, does not bring about the circumstances where additional constitutive phenomena become relevant in the flow.  For example, Coulomb theory neglects rate-sensitivity of the material, which is known to exist (i.e. the $\mu(I)$ rheology \cite{dacruz05,jop06}).  One might expect that for more rapidly spinning wheels this effect could be significant, but oftentimes modifications for rate-dependency have a minimal effect on the flow solution \cite{lagree2011granular} because the rate-effects are scaled by pressure through the inertial number.  
Size-effects and history-dependent phenomena (e.g. stress-dilatancy and anisotropy evolution) could also be less significant in locomotive intrusion processes since we deal with  macroscopic intruders (as opposed to those competing with the grain size \cite{goldsmith2013drag}) that quickly induce large strains of the surrounding media causing transients to pass rapidly. 

A more complex scaling relationship may be needed to go beyond monotonic driving, e.g. oscillatory wheel motion, where the reaction force of the material on the intruder could be more influenced by transient phenomena.  Within this caveat, the scaling could be extended to other types of \edit{locomotion, such as undulating self-propulsion}, with more moving parts by adding additional non-dimensional groups for each new degree of freedom. Testing these extended scaling laws would be important future work.

\bibliography{granular}

\end{document}